\documentstyle[11pt]{article}

\begin{document}

\centerline{\Large \bf Supersymmetry and Inflation}

\begin{center}
{{\bf G. Lazarides} \\
Physics Division, School of Technology,\\ 
University of Thessaloniki,
Thessaloniki 540 06, Greece}
\end{center}

\vspace{1.2cm} \centerline{\bf Abstract} A variant of hybrid inflation
 which is applicable in a wide class of
supersymmetric grand unified models and reproduces the observed temperature
perturbations of cosmic background radiation with natural values of the
parameters is presented. The theory is consistent with the unification of
the minimal supersymmetric standard model gauge couplings as measured at
LEP. The termination of inflation is smooth and does not produce any
topological defects. Numerical investigation of the cosmological evolution
of the system shows that for almost all initial values of the fields we do
get an adequate amount of inflation. Finally, the "reheating" process
following inflation and the production of the baryon asymmetry of the
universe via a primordial lepton asymmetry are briefly discussed and some
important implications for right handed neutrino Majorana masses are
investigated. \\

Some years ago Linde [1] has proposed, in the context of non-supersymmetric
grand unified theories (GUTs), a clever infationary scenario which he called
hybrid inflation. The idea was to use two real scalar fields $\chi $ and $
\sigma $ instead of one that is normally used. The field $\chi $ provides
the vacuum energy which drives inflation while $\sigma $ is the slowly
varying field during inflation. The main advantage of this scenario is that
it can reproduce the observed temperature fluctuations of cosmic background
radiation (CBR) with ''natural'' values of the parameters of the theory in
contrast to previous realizations of inflation which require extremely small
coupling constants. The potential utilized by Linde is

\begin{equation}
V ( \chi, \sigma)= \kappa^2 ( \mu^2 - \frac {\chi^2}{4})^2 + \frac{\lambda^2
\chi^2 \sigma^2} {4} + \frac {m^2\sigma^2}{2},
\end{equation}
where $\kappa, \lambda$ dimensionless positive coupling constants and $\mu,$
m mass parameters. The vacua lie at $<\chi>= \pm 2 \mu$, $<\sigma> =0$.
Putting m=0 for the moment, we observe that the potential possesses an exact
flat direction at $\chi=0$ with $V(\chi=0 ,\sigma)=\kappa^2 \mu^4$. The mass
squared of the field $\chi$ along this flat direction is given by $m^2_\chi
= - \kappa^2 \mu^2 + \frac{1}{2} \lambda^2 \sigma^2$ and remains 
non-negative for $\sigma \geq \sigma_c = \sqrt {2} \kappa \mu/ \lambda $. 
This means that, at $\chi=0$ and $\sigma \geq \sigma_c$, we obtain a valley
of minima with flat bottom. Reintroducing the mass parameter m in eq.(1) we
observe that this valley acquires a non-zero slope. A region of the universe
where $\chi$ and $\sigma$ happen to be almost uniform with negligible
kinetic energies and with values close to the bottom of the valley of minima
follows this valley in its subsequent evolution and undergoes inflation. The
temperature fluctuations of CBR produced during this inflation can be
estimated to be 
\begin{equation}
\frac {\delta T}{T} \simeq (\frac {32 \pi}{45})^{1/2} \frac {V^{3/2}} {M^3_P
V^\prime} \simeq (\frac {16 \pi}{45})^{1/2} \frac{\lambda \kappa^2 \mu^5}{%
M^3_Pm^2},
\end{equation}
where $M_P=1.22 \times 10^{19}$ GeV is the Planck mass, prime denotes
derivative with respect to $\sigma$ and $V,V^\prime$ are evaluated at $\chi=0
$. The Cosmic Background Explorer (COBE) result, $\delta T/T \simeq 6.6
\times 10^{-6}$, can then be reproduced with $\mu=2.86 \times 10^{16}$ GeV
[the supersymmetric (SUSY) GUT vacuum expectation value (vev)] and $m \simeq
\kappa \sqrt {\lambda}\> 1.3 \times 10^{15}$ GeV $\sim 10^{12}$ GeV for $%
\kappa, \lambda \sim 10^{-2}$. Inflation terminates abruptly at $%
\sigma=\sigma_c$ and is followed by a "waterfall", i.e., sudden entrance
into an oscillatory phase about a global minimum. Since the system can fall
into either of the two available global minima with equal probability,
topological defects can be easily produced if they are predicted by the
particular particle physics model one is considering.

The hybrid inflationary scenario is "tailor made" for application to SUSY
GUTs except that the mass of $\sigma, m, $ is unacceptably large for SUSY
where all scalar fields acquire masses of order $m_S \sim 1$ TeV from SUSY
breaking. To see this consider a SUSY GUT with a (semi-simple) gauge group $G
$ of rank $\geq 5$: 
\begin{equation}
G \to G_S \equiv SU(3)_c \times SU(2)_L \times U(1)_Y
\end{equation}
at a scale $M_X \simeq 2 \times 10^{16}$GeV. The spectrum of the theory
below $M_X$ is assumed to coincide with the minimal supersymmetric standard
model (MSSM) spectrum plus standard model (SM) singlets so that the
successful predictions for $a_s$, $sin^2 \theta_w$ are retained. The theory
may also possess global symmetries. The breaking in eq.(3) is achieved
through the superpotential 
\begin{equation}
W = \kappa s (- \mu^2 + \bar{\phi} \phi),
\end{equation}
where $\bar{\phi}, \phi$ is a conjugate pair of SM singlet left handed
superfields which belong to non-trivial representations of $G$ and reduce
its rank by their vevs and $s$ is a gauge singlet left handed superfield.
This superpotential is the most general renormalizable superpotential
consistent with a U(1) R-symmetry under which $W \to e^{i\theta} W,s \to
e^{i \theta}s, \bar {\phi} \phi \to \bar{\phi} \phi$ and gives the potential 
\begin{eqnarray*}
V=\kappa^2 \mid \mu^2 - \bar{\phi} \phi \mid^2 +\kappa^2 \mid s \mid^2 (\mid
\phi \mid^2 + \mid \bar{\phi}\mid^2)
\end{eqnarray*}
\begin{equation}
+ D-terms.
\end{equation}
Restricting ourselves to the D-flat direction $\bar{\phi}^* = \phi$ which
contains the SUSY minima and performing appropriate gauge and $R$%
-transformations we can bring $s, \bar{\phi}, \phi$ on the real axis, i.e., $%
s \equiv \sigma/ \sqrt{2}, \bar{\phi}=\phi \equiv \chi/2$, where $\sigma,
\chi$ are normalized real scalar fields. The potential then takes the form
in eq.(1) with $\kappa = \lambda$ and $m=0$ and, thus, Linde's potential for
hybrid inflation is almost obtainable from SUSY GUTs but without the mass
term of $\sigma$ which is,however, crucial since it provides the slope of
the valley of minima necessary for inflation.

One way to obtain a valley of minima useful for inflation is to replace the
renormalizable trilinear term in $W$ in eq.(4) by the next order non-
renormalizable coupling: 
\begin{equation}
W = s ( -\mu^2 +\frac {(\bar{\phi} \phi)^2}{M^2}),
\end{equation}
where $M\sim10^{18}$ GeV and is related to the scale which controls the non-
renormalizable contributions (The coupling constant $\kappa$ is absorded in $%
\mu$ and M) \cite{2}. This is achieved by imposing an extra $Z_2$ discrete
symmetry under which $\bar{\phi} \phi \to - \bar{\phi}\phi$. Indeed $W$ in
eq.(6) contains all the dominant terms consistent with this discrete, the $R-
$ and the gauge symmetry. The potential obtained for $W$ in eq.(6) is then 
\begin{equation}
V(\chi, \sigma)=(\mu^2- \frac {\chi^4}{16M^2})^2 + \frac {\chi^6
\sigma^2}{16M^4}
\end{equation}
with SUSY vacua at $<\chi>\>= \> \pm 2 (\mu M)^{1/2}$,$<\sigma>=0$. The SUSY
GUT scale is $M_X=g (\mu M)^{1/2} $ with $g \simeq 0.7$ being the unified
gauge coupling constant. The potential in eq.(7) although of quite similar
form with the potential in eq.(1) with $m=0$ possesses some crucial
differences. The flat direction at $\chi=0$ with $V(\chi=0, \sigma)= \mu^4$
is now a local maximum in the $\chi$ direction for all values of $\sigma$
but two valleys of local minima develop on both sides and close to this flat
direction at $\chi^2 \simeq 4 \mu^2 M^2/3 \sigma^2 (\sigma^2>> \mu M)$ with $%
V \simeq \mu^4(1-2 \mu^2 M^2/27 \sigma^4)$. These valleys have an inbuilt
slope and thus they can, in principle, be used for inflation. Indeed, as one
can easily show, a region of the universe with $\sigma, \chi$ field values
close to the bottom of one of these valleys follows this valley in its
subsequent evolution and inflates till $\sigma \simeq \sigma_o \simeq (2
M_P/9 \sqrt{\pi} (\mu M)^{1/2})^{1/3} (\mu M)^{1/2}$. The number of
e-foldings produced when the $\sigma$ slowly varies from an initial value $%
\sigma$ till $\sigma_o$ is estimated to be $N(\sigma) \simeq (3 \sqrt{2 \pi}
/2\mu M M_P)^2 \sigma^6$ and, consequently, the value of $\sigma$ at which
our present horizon size crossed outside the inflationary horizon is $%
\sigma_H =(9N_H/2)^{1/6} \sigma_o$, where $N_H \simeq 60$ is the number of
e-foldings suffered by our present horizon during inflation. After the end
of inflation at $\sigma_o$, the fields $\sigma, \chi$ enter smoothly into an
oscillatory phase about the SUSY vacuum with frequencies $m_\sigma = m_\chi
= 2 \sqrt{2} (\mu/M)^{1/2} \mu$. This is the reason for calling our scenario
smooth hybrid inflation.

Now, calculating the scalar part of $\delta T/T$ we find 
\begin{equation}
(\frac{\delta T} {T})_S \approx \frac {1}{\sqrt{5}} (\frac {6}{\pi})^{1/3}
N^{5/6}_H (\frac {M_X}{g})^{10/3}M^{-4/3}_P M^{-2}
\end{equation}
and the COBE result can be reproduced with $M \approx 8.6 \times 10^{17} $%
GeV, $\mu \approx 9.5 \times 10^{14}$ GeV.Consequently, the observed
temperature fluctuations of CBR can be obtained with "natural" values of the
parameters (in particular, $M \sim 10^{18}$ GeV) and consistently with the
SUSY GUT scale $M_X \approx 2 \times 10^{16}$ GeV. Some comments are in
order:(1) The gravitational wave part of $\delta T/T, (\delta T/T)_T \sim
10^{-9} <<(\delta T/T)_S$ and the spectral index $n \approx 0.97.$ (2) Since
the system follows a particular valley of minima from the beginning it ends
up at a particular SUSY vacuum and, thus, no topological defects are
produced.(3) Replacement of global SUSY with supergravity (SUGRA) makes
inflation in general impossible since it produces a mass for the inflaton
higher than the Hubble constant during inflation. In our model, due to the
R-symmetry, this does not happen so we can hope that SUGRA may not destroy
the whole picture. (4) The relevant part of inflation takes place between
the $\sigma$ field values $\sigma_H \approx 2.7 \times 10^{17}$GeV and $%
\sigma_o \approx 1.1 \times 10^{17}$ GeV which are much smaller than $M_c
\equiv M_P/ \sqrt{8 \pi} \simeq 2.4 \times 10^{18}$ GeV and, thus, SUGRA
corrections are under control.

An inflationary scenario is fully successful if it is obtainable for a wide
class of "natural" initial conditions, i.e., initial values of the fields
and their time derivatives. We, thus, tried to specify initial values of $%
\chi, \sigma$ (for simplicity we put their time derivatives equal to zero)
for which the system falls at the bottom of the valley of minima at a $%
\sigma \geq \sigma_H $ so that its subsequent evolution along the valley
produces adequate amount of inflation \cite{3} . The evolution equations are 
\begin{eqnarray*}
\ddot{\chi} +3H \dot {\chi} - \frac{\chi^3}{2M^2} (\mu^2 - \frac{\chi^4}{%
16M^2}) + \frac {3 \chi^5 \sigma^2}{8M^4} = 0,
\end{eqnarray*}
\begin{equation}
\ddot{\sigma} + 3H \dot{\sigma} + \frac{\chi^6 \sigma}{8M^4} = 0,
\end{equation}
where 
\begin{equation}
H=(\frac{8 \pi} {3})^{1/2} M_P^{-1} (\frac{1}{2} \dot{\chi} ^2 +\frac {1}{2}
\dot {\sigma}^2 +V(\sigma, \chi))^{1/2}
\end{equation}
is the Hubble constant and overdots denote time derivatives. We integrated
these equations numerically for initial values of the fields $0.1 \leq \hat{%
\sigma} \equiv \sigma/M_P \leq 1.2$, $0.01 \leq \hat{\chi} \equiv\chi /M_P
\leq 0.5$ and the results are shown in Figure 1. Each point on the $\hat{%
\sigma}- \hat{\chi}$ plane corresponds to a given set of initial conditions
and depicts a definite evolution pattern of the system. Filled circles
correspond to an evolution pattern where both $\chi$ and $\sigma$ oscillate
and fall into a SUSY vacuum without producing any appreciable amount of
inflation. Open triangles cover a region where $\sigma >> \chi$ initially
and depict a pattern where $\sigma$ decreases slowly and monotonically
towards a constant value greater than $\sigma_H$ while $\chi$ oscillates and
relaxes at the bottom of the valley. The subsequent evolution along the
valley produces adequate amount of inflation. The limiting case $\sigma>>M_P
>>\chi$ which lies deeply into the region of open triangles can be studied
analytically and one finds very little variation of $\sigma$ in this case.
Although the region of open triangles leads to adequate inflation we do not
consider it as "natural" because it requires a considerable descrepancy
between the values of $\chi$ and $\sigma$. Open circles which cover most of
the diagram correspond to the following situation: Both fields start
oscillating from the beginning but, due to energy transfer from $\chi$ to $%
\sigma$, the amplitude of $\sigma$ increases gradually and eventually the
system gets trapped to an evolution pattern of the open triangle type
leading again to adequate inflation. This interesting energy transfer
phenomenon between the two coupled oscillating fields is very important
because it ensures adequate inflation in the most "natural" region of
initial conditions $(\hat{\chi} \sim \hat {\sigma} \sim 10^{-1})$ with
"natural" initial energy density values. To avoid strong influence from
SUGRA we also considered initial field values $0.01 \leq \hat{\sigma}, \hat{%
\chi} \leq 0.1$. We found that almost all points are of the open circle
type. The overall conclusion is that almost all initial conditions give
adequate smooth hybrid inflation. This fact together with the other
advantages of this scenario makes it very "natural" and successful.

Finally we will discuss the "reheating" process \cite{4} . After the end of
inflation the inflaton, which consists of the two complex scalar fields $s$
and $\theta = (\delta \phi + \delta \bar{\phi}) /\sqrt{2}$, where $\delta
\phi = \phi - (\mu M)^{1/2}, \delta \bar{\phi} = \bar {\phi} - (\mu M)^{1/2}$%
, with mass $m_{inf} =m_s=m_\theta=2\sqrt{2}(\mu/M)^{1/2} \mu \approx 8.93
\times 10^{13}$GeV, performs damped oscillations about the SUSY vacuum and
eventually decays to light particles. We will assume that s decays faster
and, thus, we will concentrate on the decay of $\theta$. Its dominant decay
mode is in a pair of right handed neutrinos, i.e., $\theta \to \nu^c \nu^c$
through the superpotential term $\delta W= (M_{\nu^c} /2 \mu M) \bar{\phi} 
\bar{\phi} \nu^c \nu^c (M_{\nu^c}$ is the mass of the relevant right handed
neutrino). The "reheat" temperature is estimated to be $T_r \approx (1/7)(
\Gamma_\theta M_P)^{1/2} \approx 3.3 \times10^{-2} M_{\nu^c}$, where $%
\Gamma_\theta$ is the decay rate of $\theta$. For simplicity, we will ignore
the first family of quarks and leptons. The Majorana mass matrix of right
handed neutrinos can be brought to the diagonal form $m=diag(M_1, M_2)$ with 
$M_1 \geq M_2 >0$. The approximate see-saw light neutrino mass matrix $m_D 
\frac{1}{m} m_D$($m_D$ is the neutrino Dirac mass matrix) can then also be
diagonalized by rotating to an appropriate basis of left handed neutrinos.
In this basis of right and left handed neutrinos the elements of the Dirac
mass matrix 
\begin{equation}
m_D = \left[ 
\begin{array}{cc}
a & b \\ 
c & d
\end{array}
\right]
\end{equation}
are not all independent. They can be expressed in terms of three complex
parameters $a, d$ and $\eta:b=-(M_2/M_1)^{1/2} \eta a, c=(M_1/M_2)^{1/2}
\eta d $ and the light neutrino masses take the form $m_1= \mid a \mid^2$ $%
\mid 1+\eta^2 \mid /M_1, m_2=\mid d \mid^2 \mid 1+\eta^2 \mid /M_2$.
Restricting ourselves to the case $\mid \eta \mid \sim 1$ and $M_1/M_2 >>1$,
we have $\mid a \mid >> \mid b \mid , \mid c \mid >> \mid d \mid$. Taking
further $\mid c \mid << \mid a \mid, a$ becomes the dominant element of $m_D$%
. Diagonalization of $m_D^+m_D$ under these conditions gives the approximate
Dirac mass eigenvalues $\mid a\mid, \mid d \mid \mid 1+\eta^2 \mid$.
Assuming that the dominant contributions to $m_D$ and the up type quark mass
matrix coincide asymptotically we can then obtain the asymptotic relation $%
m_t \approx \mid a \mid$. We are now ready to draw some important
conclusions. Suppose that the inflaton decays predominantly to the heaviest
right handed neutrino with mass $M_1$. The well known gravitino problem
requires $T_r \leq 10^9 $GeV which in this case gives $M_1 \leq 3 \times
10^{10}$ GeV with the consequence that $m_1 \approx m^2_t \mid 1+\eta^2 \mid
/M_1 >> 100$ eV. This is cosmologically unacceptable and thus we are led to
the conclusion that the inflaton should decay to the second heaviest right
handed neutrino. This requires that $M_1 \geq m_{inf}/2 \approx
4.47\times10^{13}$ GeV and $M_2 \leq 3\times 10^{10}$ GeV. The observed
baryon asymmetry of the universe can be generated by first producing a
primordial lepton asymmetry via the out-of-equilibrium decay of the right
handed neutrinos which emerge from the decay of the inflaton.
Non-perturbative effects at the electroweak transition convert a fraction of
this asymmetry to the observed baryon asymmetry. The primordial lepton
asymmetry is given by 
\begin{equation}
\frac{n_L}{s} \approx \frac {9}{8 \pi} \frac {T_r}{m_{inf}}\frac {M_2}{M_1}
\frac {m^2_t}{\upsilon^2} Im (\eta^{*2} / \mid \eta \mid^2)
\end{equation}
($\upsilon = 174$ GeV is the electroweak scale and $m_t \approx 110 $ GeV
asymptotically) and is maximized by saturating the previously obtained
bounds on $M_1, M_2$. Infact $M_1 \approx 4.47 \times 10^{13}$ GeV and $M_2
\approx 3 \times 10^{10}$ GeV give $n_L/s \leq 10^{-9}$ which is
satisfactory. We see that the requirement of a successful "reheating"
process specifies the Majorana masses of the two heaviest right handed
neutrinos with considerable accuracy.

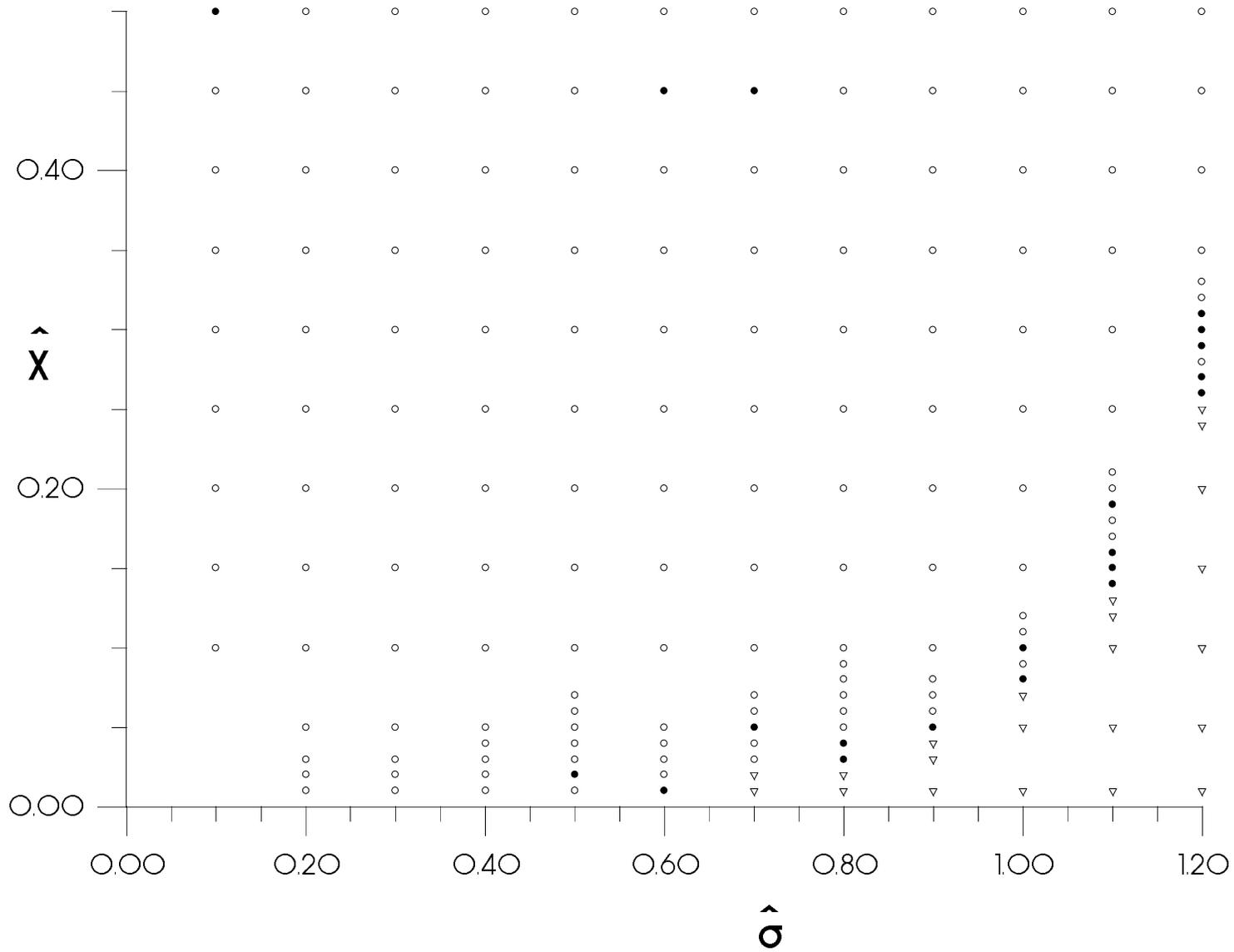
\noindent {\bf Figure 1.} Evolution patterns for the $\hat{\sigma} -\hat{\chi}$ system.

\end{document}